\documentclass[aps,prl,twocolumn,superscriptaddress]{revtex4}
\usepackage{graphicx}
\usepackage{amssymb}
\usepackage{amsmath}
\usepackage{graphicx}
\usepackage{epsfig}
\usepackage{color}
\usepackage{bm}

\begin{document}

\title{$\mathbb{Z}_2$ metallic spin liquid on a frustrated Kondo lattice}
\author{Jiangfan Wang}
\affiliation{Beijing National Laboratory for Condensed Matter Physics and Institute of Physics,
Chinese Academy of Sciences, Beijing 100190, China}
\author{Yi-feng Yang}
\email[]{yifeng@iphy.ac.cn}
\affiliation{Beijing National Laboratory for Condensed Matter Physics and Institute of Physics, 
Chinese Academy of Sciences, Beijing 100190, China}
\affiliation{School of Physical Sciences, University of Chinese Academy of Sciences, Beijing 100190, China}
\affiliation{Songshan Lake Materials Laboratory, Dongguan, Guangdong 523808, China}
\date{\today}

\begin{abstract}
Metallic spin liquid has been reported in several correlated metals, but a satisfactory theoretical description is not yet available. Here we propose a potential route to realize the metallic spin liquid and construct an effective $\mathbb{Z}_2$ gauge theory with charged fractionalized excitations on the triangular Kondo lattice. This leads to a $\mathbb{Z}_2$ metallic spin liquid featured with long-lived, heavy holon excitations of spin 0 and charge $+e$ and a partially enlarged electron Fermi surface. It differs from the weak-coupling FL$^*$ state proposed earlier and may be viewed as a \textit{fractionalized heavy fermion liquid}. Our theory provides a general framework to describe the metallic spin liquid in frustrated Kondo lattice systems.
\end{abstract}

\maketitle

Quantum spin liquid (QSL), characterized by topological orders and fractionalized excitations, is usually expected in frustrated spin systems and has attracted tremendous interest in past decades \cite{Balents2010,Broholm2020}. Its counterpart in metals, dubbed the \textit{metallic spin liquid}, is less explored despite that experimental signatures have been found in, e.g., the heavy fermion compounds CePdAl \cite{zhao2019CePdAl, zhangPRB2018}, Pr$_2$Ir$_2$O$_7$ \cite{Nakatsuji2006, Tokiwa2014}, LiV$_2$O$_4$ \cite{Okabe2019}, and the organic material $\kappa$-(ET)$_4$Hg$_{2.89}$Br$_8$ \cite{Oike2017}. A candidate description of the metallic spin liquid has been proposed previously on Kondo lattices and given the name fractionalized Fermi liquid (FL$^*$), where the QSL is weakly coupled to conduction electrons \cite{Senthil2003prl,Senthil2004prb}. As a result, the Fermi surface only contains conduction electrons (small) and the spin liquid part is essentially untouched \cite{Senthil2003prl,Senthil2004prb,Ashivin2004Luttinger}. It is in sharp contrast to the heavy Fermi liquid (HFL) in the strong-coupling limit, where the Fermi surface is large and contains both conduction electrons and local spins due to the Kondo screening. This raises the question concerning how the FL$^*$ and HFL states are connected with increasing Kondo coupling \cite{si2006global,coleman2010frustration,Wang2021nonlocal} and if there may exist a metallic spin liquid beyond such a weak-coupling description. 

Some insights may be borrowed from the cuprates \cite{Sachdev2012}. In a slightly doped Mott insulator, it has been proposed \cite{Kivelson1987Soliton} that a conduction hole can combine with a spinon to form a spinless charged particle called holon \cite{Anderson1988}, which is physically related to the Zhang-Rice singlet formed due to the Kondo coupling between doped holes on the oxygen $p$ orbitals and the copper spins \cite{ZhangRice1988}. Ideas based on such spin-charge fractionalization have been extensively investigated in theory \cite{PALee_RMP2006, SenthilFisher2000,Kaul2007ACL} and explored in experiment  \cite{Graf2007,Doiron2007,Ayres2021}. In the Kondo lattice, holon may also be formed as a quasi-bound state of bosonic spinon and conduction hole \cite{pepin2005}. A recent large-$N$ mean-field calculation predicted that the holons can have a dispersion once nonlocal spatial correlations are correctly included \cite{Wang2021nonlocal}. This leads to an intermediate holon state connecting the weak-coupling FL$^*$ and the strong-coupling HFL, but it is not clear if holons can survive against gauge fluctuations. 

In this work, we elaborate this idea by considering the simplest gapped two-dimensional QSL, the $\mathbb{Z}_2$ short-ranged resonating valence bond state, and develop an effective gauge theory of the holons on a triangular Kondo lattice. We go beyond the mean-field calculations and demonstrate the holon stability against  $\mathbb{Z}_2$ gauge fluctuations. Our calculations reveal a $\mathbb{Z}_2$ metallic spin liquid with mobile, heavy, long-lived holon excitations and a partially enlarged electron Fermi surface, which may be viewed as a \textit{fractionalized heavy fermion liquid} (FHF) differing from the usual QSL and the weak-coupling FL$^*$. Our method provides a general framework to study metallic spin liquid on frustrated Kondo lattices.

We start with the Kondo-Heisenberg model on a triangular lattice:
\begin{eqnarray}
H=t\sum_{\left\langle ij\right\rangle \alpha a}c_{i\alpha a}^\dagger c_{j\alpha a}+J_K\sum_{i}{\bf S}_{i}\cdot {\bf s}_{i}+J_H\sum_{\left\langle ij\right\rangle}{\bf S}_{i}\cdot {\bf S}_{j},
\label{eq:H}
\end{eqnarray}
where the conduction electron $c_{i\alpha a}$ has a spin index ($\alpha$) and a channel index ($a=1,\cdots, K$), ${\bf s}_{i}$ is its spin density, and ${\bf S}_{i}$ describes the local spin. $J_K$ and $J_H$ are the antiferromagnetic Kondo and Heisenberg coupling constants, respectively. The Schwinger boson representation states ${\bf S}_i=\frac{1}{2}\sum_{\alpha\beta}b_{i\alpha}^\dagger \bm{\sigma}_{\alpha\beta}b_{i\beta}$, where $b_{i\alpha}^\dagger$ creates a bosonic spinon with the local constraint $n_i^b\equiv \sum_\alpha b_{i\alpha}^\dagger b_{i\alpha}=2S$ \cite{Arovas1988}. The Kondo screening is perfect at large $J_K$ by choosing $K=2S$ \cite{parcollet1997transition}. 

The $\mathbb{Z}_2$ QSL is the ground state of the Sp($N$) extension of the Heisenberg Hamiltonian at small $\kappa\equiv 2S/N$ denoting strong quantum fluctuations \cite{Read1992Kagome}. It can be obtained by decomposing the Heisenberg term, $J_H{\bf S}_i\cdot{\bf S}_j \rightarrow\sum_{\alpha}(\Delta_{ij}\text{sgn}(\alpha) b_{i\alpha}b_{j,-\alpha}-\Gamma_{ij}b_{i\alpha}^\dagger b_{j\alpha})+c.c.+2N(|\Delta_{ij}|^2-|\Gamma_{ij}|^2)/J_H$, where $\alpha=\pm 1,\cdots, \pm N$ is the spin index, and $\Delta_{ij}$ and $\Gamma_{ij}$ are two auxiliary fields describing the spinon pairing and hopping amplitudes, respectively \cite{FlintSpN2009,WangFa2006,Wang2022_3IK}. The local constraint is imposed by a Lagrange multiplier $\lambda_i$. The fluctuations of these fields introduce an U(1) gauge field coupled to the spinons \cite{Read1992Kagome}. On a triangular lattice, the spinon pair condensate reduces U(1) to $\mathbb{Z}_2$ at low energy via the Higgs mechanism \cite{Read1992Kagome}. 

The holons are described by a fermionic auxiliary field $\chi_{ia}$ and emerge from the Kondo coupling between spinons and conduction electrons via a Hubbard-Stratonovich transformation, $J_K{\bf S}_i\cdot {\bf s}_i\rightarrow\frac{1}{\sqrt{N}}\sum_{\alpha a}b_{i\alpha}^\dagger c_{ia\alpha}\chi_{ia}+c.c.+\sum_a |\chi_{ia}|^2/J_K$ \cite{Wang2021nonlocal, Coleman-SWB-2006, Yashar1D,Komijani2019, Wang2019quantum, Wang2022_3IK, Wang2022_FM, Han2021}. We have eventually an interacting system consisting of spinons, holons, and conduction electrons \cite{supp}:
\begin{equation}
\mathcal{L}=\mathcal{L}_{c}+\mathcal{L}_{b}+\mathcal{L}_{\chi}+\frac{1}{\sqrt{N}}\sum_{i\alpha a}\left(b_{i\alpha}^\dagger c_{ia\alpha}\chi_{ia}+c.c.\right),
\end{equation}
where $\mathcal{L}_{c}=\sum_{{\bf k}\alpha a}c_{{\bf k}\alpha a}^\dagger (\partial_\tau+\epsilon_{\bf k})c_{{\bf k}\alpha a}$ gives the electron dispersion, $\mathcal{L}_\chi=\sum_{ia}|\chi_{ia}|^2/J_K$ is the holon action, and $\mathcal{L}_b$ describes the $\mathbb{Z}_2$ QSL with bosonic spinons.
 
\textit{The large-$N$ solution.---} We first focus on the uniform mean-field solution assuming $\Delta_{i,i+\eta}=\Delta_0$, $\Gamma_{i,i+\eta}=\Gamma_0$, and $\lambda_i=\lambda_0$, where $\eta=(1,0), (-\frac{1}{2},\pm\frac{\sqrt{3}}{2})$ are three unit vectors. This describes the ``zero-flux'' state of the triangular lattice Heisenberg model at small $\kappa$ \cite{WangFa2006}. Upon increasing $\kappa$, the spinons condense on the corner points of the hexagonal Brillouin zone, leading to the 120$^\circ$ N\'eel order \cite{WangFa2006,Read1992Kagome}. The electron-spinon-holon vertex leads to the following self-energy equations \cite{Wang2021nonlocal,Wang2022_FM}:
\begin{eqnarray}
\Sigma _{b}({\textbf p}, i\nu_n)&=&-\frac{\kappa}{\beta\mathcal{V}} \sum_{{\textbf k}l}g_c({\textbf p}-{\textbf k}, i\nu_n-i\omega_l ) G_{\chi }({\textbf k}, i\omega_l ), \notag \\
\Sigma _{\chi }({\textbf p}, i\omega_n)&=&\frac{1}{\beta\mathcal{V}}\sum_{{\textbf k}l}g_c({\textbf k}-{\textbf p}, i\nu_l -i\omega_n ) G_{b}({\textbf k}, i\nu_l ),
\label{eq:SelfE}
\end{eqnarray}
where $g_c$ is the bare Green's function of conduction electrons, $G_{b}$ and $G_{\chi }$ are the full Green's functions of spinons and holons. The self-energy of conduction electrons is proportional to $1/N$ and vanishes in the large-$N$ limit. Using fast Fourier transform, we are able to solve the above self-consistent equations efficiently in real space \cite{supp}, and predict a dispersive holon band that is impossible in the local approximation \cite{Yashar1D,Komijani2019,Wang2019quantum}, which is the key for the occurrence of the metallic spin liquid.

\begin{figure}
\centering\includegraphics[scale=0.37]{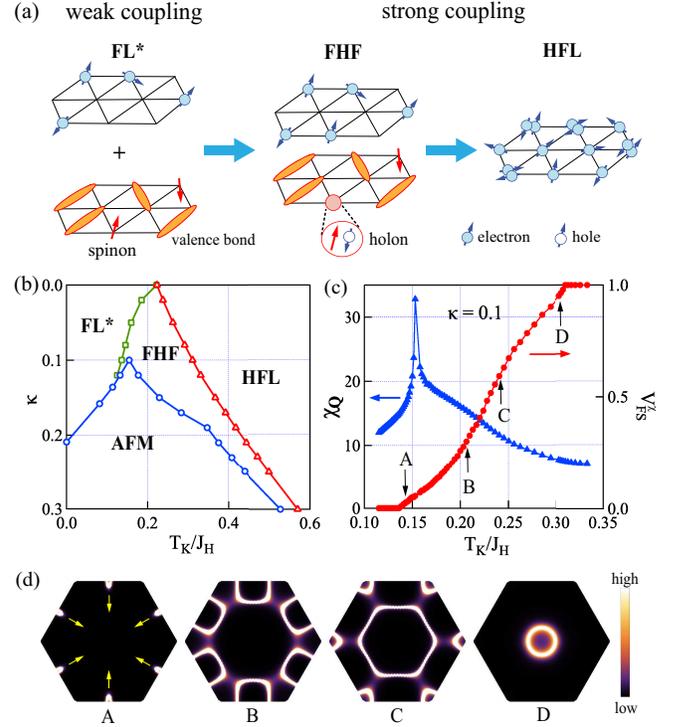}
\caption{(a) Schematic diagrams showing different paramagnetic phases on a frustrated Kondo lattice upon increasing Kondo coupling between conduction electrons (top layer) and local spins (bottom layer). From left to right: the weak-coupling fractionalized Fermi liquid (FL$^*$), the strong-coupling fractionalized heavy fermion (FHF) liquid and the heavy Fermi liquid (HFL). (b) The large-$N$ phase diagram of the triangular lattice Kondo-Heisenberg model. AFM denotes the 120$^\circ$ N\'eel order. (c) Spin susceptibility at the ordering wave vector $\mathbf{Q}=\pm (\frac{2\pi}{3}, \frac{2\pi}{\sqrt{3}})$, and the holon Fermi volume $V_{FS}^\chi$ as functions of $T_K/J_H$ for $\kappa=0.1$. (d) The holon Fermi surfaces in the FHF state at different $T_K/J_H$ marked by $A$-$D$ in (c).  The yellow arrows mark the direction of expansion of the holon Fermi pockets with increasing $T_K/J_H$.}
\label{fig:Global}
\end{figure}

\textit{The global phase diagram.---} Figure \ref{fig:Global}(b) shows the large-$N$ zero temperature phase diagram in terms of $\kappa$ and $T_K/J_H$, where $T_K$ is the single ion Kondo temperature. There are three transition lines separating four phases: the fractionalized Fermi liquid (FL$^*$), the fractionalized heavy fermion liquid (FHF), the heavy Fermi liquid (HFL), and the antiferromagnetic (AFM) state with an ordering wave vector ${\bf Q}=\pm(\frac{2\pi}{3},\frac{2\pi}{\sqrt{3}})$ corresponding to the 120$^\circ$ N\'eel order. The FL$^*$ and HFL states can be understood from the Heisenberg and the Kondo limits, respectively. The former has a small electron Fermi surface weakly coupled to a $\mathbb{Z}_2$ QSL \cite{Senthil2003prl}, while the latter has only electron excitations with a large Fermi surface, as illustrated in Fig. \ref{fig:Global}(a).

For small $\kappa$, these two states are separated by an intermediate state where spinons combine with conduction holes to form heavy holons with fermionic statistics, as shown in Fig.  \ref{fig:Global}(a). An alternative view (from the large-$U$ Anderson lattice) is that the renormalized heavy $f$-holes are fractionalized into bosonic spinons and fermionic heavy holons, hence the name \textit{fractionalized heavy fermion liquid}. Both spinons and holons are minimally coupled to a $ \mathbb{Z}_2$ gauge field, and are propagating particles due to the deconfinement of the gauge theory. The dispersive holon band allows us to define a gauge-invariant Fermi volume, $V_{FS}^\chi=\frac{1}{\mathcal{V}}\sum_{\bf k}\theta(-J_K^*({\bf k})^{-1})$, where $J_K^*({\bf k})=[1/J_K+\text{Re}\Sigma_\chi({\bf k},0)]^{-1}$ is the renormalized Kondo coupling inversely proportional to the holon dispersion. Upon increasing $T_K/J_H$, the holon band evolves from above to below the Fermi energy \cite{supp}, accounting for the stability of Kondo singlet formation. Its finite bandwidth inevitably leads to $0<V_{FS}^\chi<1$ and the holon Fermi surface as shown in Fig. \ref{fig:Global}(d). A generalized Luttinger sum rule requires that the electrons have a partially enlarged Fermi surface with $NV_{FS}^c=n_c+V_{FS}^\chi$ \cite{coleman2005sum}. The FL$^*$-FHF and FHF-HFL transition lines are determined by the deviation of the holon Fermi volume from $0$ and $1$, respectively, as denoted in Fig. \ref{fig:Global}(c). 

The partial enlargement of electron Fermi surface and the deconfined spinons and holons indicate that the Kondo effect exists in an incomplete and nonlocal fashion, described by the scattering process $J_K^*({\bf r}_j-{\bf r}_i)c_{ja\beta}^\dagger b_{j\beta}b_{i\alpha}^\dagger c_{ia\alpha}$ \cite{Wang2021nonlocal,Wang2022_3IK}. Intuitively, one can view the holon $\chi_{ia}^\dagger \sim \sum_\alpha b_{i\alpha}^\dagger c_{ia\alpha}$ as half a Kondo singlet, whose presence also requires the fractionalization of magnons ($S_{+}\rightarrow b_{j\uparrow}^\dagger b_{i\downarrow}$). In a three-impurity Kondo model, we have assigned such nonlocal Kondo effect to the term $c_{i\alpha}^\dagger \bm{\sigma}_{\alpha\beta} c_{j\beta} \cdot ( {\bf S}_i\times {\bf S}_j) $ \cite{Wang2022_3IK}, which emerges under the renormalization group flow and becomes strongly enhanced in certain intermediate parameter region. Similarly, the FHF phase also occurs here within an intermediate range of $T_K/J_H$, bridging the weak-coupling FL$^*$ and the strong-coupling HFL as illustrated in Fig. \ref{fig:Global}(a).

\begin{figure}
\centering\includegraphics[scale=0.3]{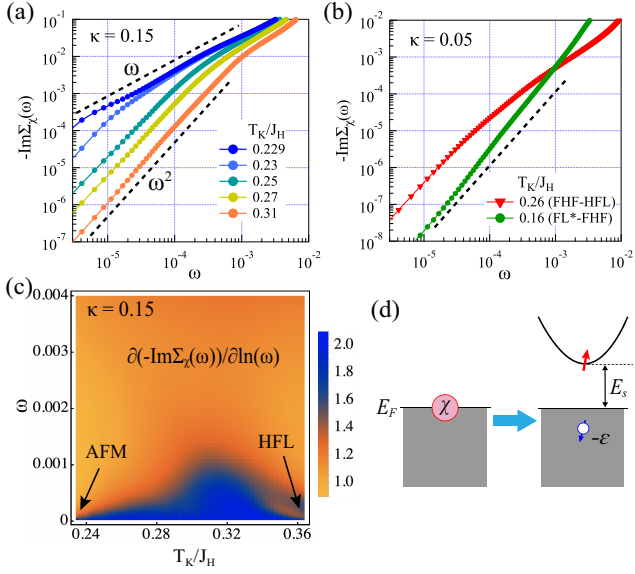}
\caption{(a) A log-log plot of the holon self-energy $-\text{Im}\Sigma_\chi(\omega)$ at $\kappa=0.15$ showing linear-in-$\omega$ behavior at the AFM QCP ($T_K/J_H=0.229$) and $\omega^2$ behavior inside the FHF ($T_K/J_H=0.23\sim 0.31$). (b) The $\omega$ dependence of $-\text{Im}\Sigma_\chi(\omega)$ at the FHF-HFL and FL$^*$-FHF transition points for $\kappa=0.05$. The dashed line denotes the $\omega^2$ behavior. (c) A color-coded plot of $\partial \ln(-\text{Im}\Sigma_\chi(\omega))/\partial \ln\omega$ on the $\omega \sim T_K/J_H$ plane for $\kappa=0.15$. (d) An illustration of the holon stability within the FHF phase: a holon at the Fermi energy ($E_F$) cannot decay into a conduction hole and a gapped spinon at low temperature. }
\label{fig:NFL}
\end{figure}

\textit{Holon dissipation.---}The holons may dissipate through the decay process $\chi_{ia}^\dagger \rightarrow c_{i\alpha a} +b_{i\alpha}^\dagger$ as reflected in its self-energy. Figure \ref{fig:NFL}(a) shows a log-log plot of $-\text{Im}\Sigma_\chi(\omega)$ at small frequency for $\kappa=0.15$, where $\Sigma_\chi(\omega)=\frac{1}{\mathcal{V}}\sum_{\bf k}\Sigma_\chi({\bf k},\omega)$ is the momentum averaged self-energy. We found $-\text{Im}\Sigma_\chi(\omega)\propto \omega^p$ with $p\approx 1$ at the AFM boundary, but $p=2$ at low frequencies deep inside the FHF phase. Approaching the FL$^*$ or HFL phase boundaries, $p$ also deviates from $2$, as shown in Fig. \ref{fig:NFL}(b) for $\kappa=0.05$, possibly associated with the Lifshitz-type transitions of the holon band at these two boundaries. Figure \ref{fig:NFL}(c) is the color-coded plot of $\partial \ln(-\text{Im}\Sigma_\chi(\omega))/\partial \ln\omega$ on the $\omega\sim T_K/J_H$ plane for $\kappa=0.15$. A clear dome of $p=2$ suggests that holons are well-defined (long-lived) quasiparticles at low energy scales in the FHF. At the perturbative level, this can be understood from the absence of phase space for holons to decay into conduction holes and gapped spinons, as illustrated in Fig. \ref{fig:NFL}(d). To be more precise, our full self-consistent calculations reveal highly damped spinons with a pseudogapped density of states within the FHF phase \cite{supp}, $\rho_b(\omega)\propto \omega$, which then indicates $-\text{Im}\Sigma_\chi(\omega)\approx \pi\rho_{c}\int_0^\omega dz \rho_b(z)\propto \omega^2$ where $\rho_{c}$ is the electron density of states at the Fermi energy. The FHF phase is therefore a metallic spin liquid featured with spin-charge fractionalization and scattering between electrons, holons, and damped spinons. At the AFM quantum critical point (QCP), the spinon gap vanishes, resulting in strong holon dissipation and the $p\approx 1$ strange metal behavior, a typical phenomenon when Fermi surfaces are coupled to critical bosonic modes associated with some magnetic instability \cite{LohneysenRMP2007}.

\textit{$\mathbb{Z}_2$ metallic spin liquid.---} The $\mathbb{Z}_2$ gauge fluctuations are associated with the dynamical sign changes of the auxiliary fields, $\Delta_{ij}=\Delta_0 Z_{ij}$, $\Gamma_{ij}=\Gamma_0Z_{ij}$, with $Z_{ij}=\pm1$ being the $\mathbb{Z}_2$ gauge field. The model is invariant under the gauge transformation $b_{i\alpha}\rightarrow b_{i\alpha}\sigma_i$, $\chi_{ia}\rightarrow \chi_{ia}\sigma_i$, $Z_{ij}\rightarrow Z_{ij}\sigma_i\sigma_j$, with $\sigma_i=\pm 1$. For small $\kappa$ above the AFM phase, the mean-field solution always satisfies $\lambda_0\gg (\Delta_0, \Gamma_0)$, as can be seen from the $\kappa=0$ limit where the spinons become exactly local ($\Delta_0=\Gamma_0=0$) \cite{Wang2021nonlocal}. This suggests a perturbative expansion in terms of $\Delta_0$ and $\Gamma_0$, which can be done by perturbatively integrating out the spinons with a nonzero gap $\lambda_0$.  Since we are mostly interested in the holon dynamics under the gauge fluctuations,  it is also helpful to integrate out the conduction electrons. At zero temperature, this procedure does not lead to divergences, reflecting the stability of holons \cite{supp}.

\begin{figure}
\centering\includegraphics[scale=0.4]{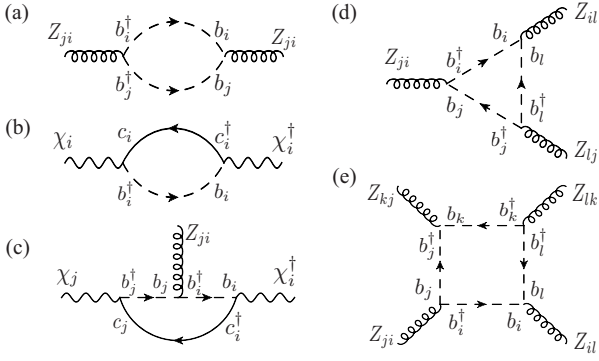}
\caption{Feynman diagrams giving rise to the effective Hamiltonian Eq. (\ref{eq:HZ2}) of the $\mathbb{Z}_2$ metallic spin liquid.}
\label{fig:Diag}
\end{figure}

The most important Feynman diagrams are listed in Fig. \ref{fig:Diag}, which give rise to the following effective Hamiltonian \cite{supp}:
\begin{eqnarray}
H_{\text{eff}}&=&-g\sum_{\langle ij\rangle}\hat{X}_{ij}+\epsilon_\chi \sum_{ia}\chi_{ia}^\dagger \chi_{ia}+\bar{t}\sum_{\langle ij\rangle a}\hat{Z}_{ij}\chi_{ia}^\dagger\chi_{ja} \notag \\
&&-K\sum_{\triangle}\prod_{ij\in\triangle}\hat{Z}_{ij}-K'\sum_{\lozenge}\prod_{ij\in\lozenge} \hat{Z}_{ij}+\cdots
\label{eq:HZ2}
\end{eqnarray}
where $\hat{X}_{ij}$ ($\hat{Z}_{ij}$) is the Pauli matrix of $x$ ($z$) component. We have scaled the holon field by its quasiparticle residue, so that $\chi_{ia}$ in Eq. (\ref{eq:HZ2}) is a canonical fermion operator.  The term $-g\hat{X}_{ij}$ comes from the dynamical term $(\partial_\tau Z_{ij})^2$ in the path integral generated by the diagram Fig. \ref{fig:Diag}(a), and tends to flip the $\hat{Z}_{ij}$ fields \cite{Kogut1979RMP}. The second and third terms are generated by Figs. \ref{fig:Diag}(b) and \ref{fig:Diag}(c), respectively. The second line of Eq. (\ref{eq:HZ2}) contains all possible interaction terms of the form $\hat{Z}_{ij}\hat{Z}_{jk}\cdots \hat{Z}_{li}$, where the links $(ij,\cdots, li)$ form a closed loop. They are generated by diagrams like Figs. \ref{fig:Diag}(d) and \ref{fig:Diag}(e), so that larger loops are associated with higher powers of $\Delta_0$ and $\Gamma_0$ and can be safely neglected. The parameters in Eq. (\ref{eq:HZ2}) are related to the microscopic parameters through:
\begin{eqnarray}
g&=&\frac{1}{\epsilon}e^{-N\Delta_0^2/(\epsilon \lambda_0^3)},\quad \epsilon_\chi\propto \frac{1}{J_K}-\rho_{c}\ln\frac{D}{\lambda_0},\quad \bar{t}\propto \Gamma_0, \notag \\
K&=&-2N\frac{\Gamma_0\Delta_0^2}{\lambda_0^2},\quad K'=N\frac{1}{\lambda_0^3}\Delta_0^2(\Delta_0^2-4\Gamma_0^2),
\label{eq:par}
\end{eqnarray}
where $\epsilon$ is the small discrete time slice of the path integral determined by the high energy cutoff and $D$ is the conduction bandwidth. The onsite energy of holons ($\epsilon_\chi$) decreases monotonically with increasing $J_K$, and becomes negative at $J_K\geq \rho_c\ln\frac{\lambda_0}{D}$, or equivalently, $T_K\geq \lambda_0$. This means the energy gain of forming a Kondo bound state ($T_K$) overcomes its least energy cost (the spinon gap $\lambda_0$). The holon bandwidth is proportional to $\Gamma_0$, which also monotonically decreases with increasing $T_K/J_H$ and eventually vanishes deep inside the HFL.

\begin{figure}
\centering\includegraphics[scale=0.42]{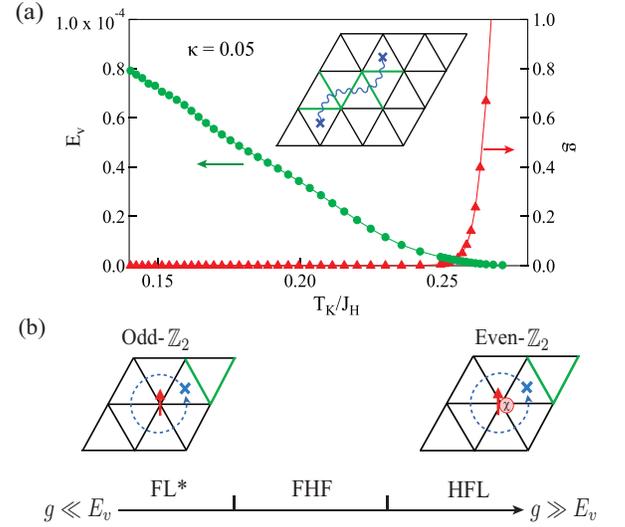}
\caption{(a) Comparison of $E_v=4K+12K'$ and $g$ as functions of $T_K/J_H$. Their values are obtained using Eq. (\ref{eq:par}) with $N=2$ and $\epsilon=0.1$. The inset shows a configuration with a pair of separated visons, where $Z_{ij}=-1$ ($1$) on the green (black) bonds and the blue ``$\times$'' represents the vison cores. (b) Different paramagnetic phases upon tuning $T_K/J_H$, or equivalently, $g/E_v$. For $2S=K=1$, the number of background $\mathbb{Z}_2$ charges at each site is $1$ in the FL$^*$ phase and  $2$ in the HFL phase, corresponding to the odd- and even-$\mathbb{Z}_2$ gauge theories, respectively. }
\label{fig:Z2}
\end{figure}

The confinement-deconfinement transition of the $\mathbb{Z}_2$ gauge theory is determined by the competition between the energy gain of bond flips ($g$) and the energy cost of producing vison excitations  with $\prod_{ij\in \mathcal{C}}Z_{ij}=-1$, where $\mathcal{C}$ is any closed loop enclosing a vison core \cite{SenthilFisher2000,Sachdev2011vison}. Ignoring the matter fields $\chi_{ia}$, creating a pair of well separated visons costs an amount of energy $E_v=4K+12K'$, which decreases with increasing $T_K/J_H$ as shown in Fig. \ref{fig:Z2}(a) for $\kappa=0.05$ by substituting our mean-field solutions into Eq. (\ref{eq:par}). By contrast, $g$ is exponentially suppressed at small $T_K/J_H$ due to the relatively large $\Delta_0$, but becomes large ($\sim\epsilon^{-1}$) due to the vanishingly small $\Delta_0$ inside the HFL. For $E_v\gg g$, the visons are expelled and the ground state can be well described by the mean-field solution with finite $\Delta_0$ and $\Gamma_0$. This is the deconfined phase with propagating fractional excitations. In the opposite limit $E_v\ll g$, visons proliferate (condense) and $\hat{Z}_{ij}$ fluctuates strongly, leading to the confinement of spinons and holons \cite{Kogut1979RMP,SenthilFisher2000}.

An important difference of the FHF metallic spin liquid from a neutral QSL is that the background $\mathbb{Z}_2$ charge density varies upon tuning $T_K/J_H$. The number of $\mathbb{Z}_2$ charges at each site is $n_i^b=2S$ in the FL$^*$ phase, but becomes $n_i^b+n_i^\chi=2S+K$ in the HFL state due to the full occupation of holon band. This introduces different Berry phases as a vison adiabatically moves around a site $i$ and returns to its original position as illustrated in Fig. \ref{fig:Z2}(b) \cite{Sachdev2011vison}. The final state of such a process is different from the initial one by a gauge transformation, $b_{i\alpha}\rightarrow -b_{i\alpha}$, $\chi_{ia}\rightarrow -\chi_{ia}$, $Z_{ij}\rightarrow -Z_{ij}$ for $j\in \text{NN}(i)$, where NN$(i)$ stands for the nearest neighbored sites of $i$. This leads to a Berry phase factor $G_i=(-1)^{n_i^b+n_i^\chi}$.  For $S=1/2$ and $K=1$, one has $G_i=-1$ in the FL$^*$ phase but $G_i=1$ in the HFL phase, corresponding to the odd- and even-$\mathbb{Z}_2$ gauge theories, respectively \cite{SenthilFisher2000,Moessner2001,Gazit2017}. We thus identify the HFL as the confined phase of an even-$\mathbb{Z}_2$ gauge theory.  Upon increasing $T_K/J_H$, the transition from the odd-$\mathbb{Z}_2$ to the even-$\mathbb{Z}_2$ theory can happen either through a single QCP with a Fermi surface jump as in the $\kappa=0$ limit, or an intermediate phase with non-integer holon filling.

The FHF metallic spin liquid shares some similarities with the algebraic charge liquid proposed earlier for the pseudogap region of cuprates \cite{Kaul2007ACL}. The latter also has fermionic holon excitations but coupled to a U(1) gauge field. The holon Fermi surface may not be directly measured via the angle-resolved photoemission spectroscopy (ARPES), but its gauge-invariant Fermi volume should in principle contribute to quantum oscillation or Hall measurements \cite{Kaul2007ACL}. Like the algebraic charge liquid, the FHF could be a parent state for other instabilities such as holon superconductivity or holon charge density wave. The former may provide additional pairing channel for heavy fermion superconductivity, while the latter breaks the translational symmetry and may be identified as the partial Kondo screening phase studied in Refs. \cite{Motome2010PKS, Assaad2015DMFT}. More investigations are needed to elaborate these possibilities.

This work was supported by the National Key Research and Development Program of China (Grant No. 2017YFA0303103), the National Natural Science Foundation of China (Grants No. 12174429, No. 11974397), and the Strategic Priority Research Program of the Chinese Academy of Sciences (Grant No. XDB33010100).


\begin{thebibliography}{100}

\bibitem{Balents2010} L. Balents, Spin liquids in frustrated magnets. Nature, {\bf 464}, 199 (2010).

\bibitem{Broholm2020} C. Broholm, R. J. Cava, S. A. Kivelson, D. G. Nocera, M. R. Norman, and T. Senthil, Quantum spin liquids. Science, {\bf 367}, 263 (2020).

\bibitem{zhao2019CePdAl} H. Zhao, J. Zhang, M. Lyu, S. Bachus, Y. Tokiwa, P. Gegenwart, S. Zhang, J. Cheng, Y.-F. Yang, G. Chen, Y. Isikawa, Q. Si, F. Steglich, and P. Sun, Quantum-critical phase from frustrated magnetism in a strongly correlated metal. Nat. Phys. {\bf 15}, 1261 (2019).  

\bibitem{zhangPRB2018} J. Zhang, H. Zhao, M. Lv, S. Hu, Y. Isikawa, Y.-F. Yang, Q. Si, F. Steglich, and P. Sun, Kondo destruction in a quantum paramagnet with magnetic frustration. Phys. Rev. B {\bf 97}, 235117 (2018).


\bibitem{Nakatsuji2006} S. Nakatsuji, Y. Machida, Y. Maeno, T. Tayama, T. Sakakibara, J. van Duijn,  L. Balicas, J. N. Millican, R. T. Macaluso, and J. Y. Chan, Metallic spin-liquid behavior of the geometrically frustrated Kondo lattice Pr$_2$Ir$_2$O$_7$. Phys. Rev. Lett. {\bf 96}, 087204 (2006).

\bibitem{Tokiwa2014} Y. Tokiwa, J. J. Ishikawa, S. Nakatsuji, and  P. Gegenwart, Quantum criticality in a metallic spin liquid. Nat. Mater. {\bf 13}, 356 (2014).

\bibitem{Okabe2019} H. Okabe, M. Hiraishi, A. Koda, K. M. Kojima, S. Takeshita, I. Yamauchi, Y. Matsushita, Y. Kuramoto, and R. Kadono, Metallic spin-liquid-like behavior of LiV$_2$O$_4$. Phys. Rev. B {\bf 99}, 041113(R) (2019).

\bibitem{Oike2017}  H. Oike, Y. Suzuki, H. Taniguchi, Y. Seki, K. Miyagawa, and K. Kanoda, Anomalous metallic behaviour in the doped spin liquid candidate $\kappa$-(ET)$_4$Hg$_{2.89}$Br$_8$. Nat. Commun. {\bf 8}, 756 (2017).

\bibitem{Senthil2003prl} T. Senthil, S. Sachdev, and M. Vojta, Fractionalized Fermi liquids. Phys. Rev. Lett. {\bf 90}, 216403 (2003).

\bibitem{Senthil2004prb} T. Senthil, M. Vojta, and S. Sachdev, Weak magnetism and non-Fermi liquids near heavy-fermion critical points. Phys. Rev. B {\bf 69}, 035111 (2004).

\bibitem{Ashivin2004Luttinger} A. Paramekanti and A. Vishwanath, Extending Luttinger's theorem to Z$_2$ fractionalized phases of matter. Phys. Rev. B {\bf 70}, 245118 (2004).

\bibitem{Wang2021nonlocal} J. Wang and Y.-F. Yang, Nonlocal Kondo effect and quantum critical phase in heavy-fermion metals. Phys. Rev. B {\bf 104}, 165120 (2021).


\bibitem{si2006global} Q. Si, Global magnetic phase diagram and local quantum criticality in heavy fermion metals. Physica B Condens. Matter {\bf 378}, 23 (2006).

\bibitem{coleman2010frustration} P. Coleman and A. H. Nevidomskyy, Frustration and the Kondo effect in heavy fermion materials. J. Low. Temp. Phys. {\bf 161}, 182 (2010). 

\bibitem{Sachdev2012} S. Sachdev, M. A. Metlitski, and M. Punk, Antiferromagnetism in metals: from the cuprate superconductors to the heavy fermion materials. J. Phys. Condens. Matter {\bf 24}, 294205 (2012).

\bibitem{Kivelson1987Soliton} S. A. Kivelson, D. S. Rokhsar, and J. P. Sethna, Topology of the resonating valence-bond state: Solitons and high-T$_c$ superconductivity. Phys. Rev. B {\bf 35}, 8865 (1987).

\bibitem{Anderson1988} P. W. Anderson and Z. Zou, ``Normal'' tunneling and ``normal'' transport: Diagnostics for the resonating-valence-bond state. Phys. Rev. Lett. {\bf 60}, 132 (1988).

\bibitem{ZhangRice1988} F. C. Zhang and T. M. Rice, Effective Hamiltonian for the superconducting Cu oxides. Phys. Rev. B {\bf 37}, 3759 (1988).

\bibitem{PALee_RMP2006} P. A. Lee, N. Nagaosa, and X.-G. Wen, Doping a Mott insulator: Physics of high-temperature superconductivity. Rev. Mod. Phys. {\bf 78}, 17 (2006).

\bibitem{SenthilFisher2000} T. Senthil and M. P. A. Fisher, Z$_2$ gauge theory of electron fractionalization in strongly correlated systems. Phys. Rev. B {\bf 62}, 7850 (2000).

\bibitem{Kaul2007ACL} R. K. Kaul, Y. B. Kim, S. Sachdev, and T. Senthil, Algebraic charge liquids. Nat. Phys. {\bf 4}, 28 (2008).

\bibitem{Graf2007} J. Graf, G.-H. Gweon, K. McElroy, S. Y. Zhou, C.  Jozwiak, E. Rotenberg,  A. Bill, T. Sasagawa,  H. Eisaki, S. Uchida, H. Takagi, D.-H. Lee, and A. Lanzara, Universal high energy anomaly in the angle-resolved photoemission spectra of high temperature superconductors: possible evidence of spinon and holon branches. Phys. Rev. Lett. {\bf 98}, 067004 (2007).

\bibitem{Doiron2007} N. Doiron-Leyraud, C. Proust, D. LeBoeuf, J. Levallois, J. Bonnemaison, R. Liang, D. A. Bonn, W. N. Hardy, and L. Taillefer, Quantum oscillations and the Fermi surface in an underdoped high-$T_c$ superconductor. Nature, {\bf 447}, 565 (2007).

\bibitem{Ayres2021} J. Ayres, M. Berben, M. Čulo, Y.-T. Hsu, E. van Heumen, Y. Huang, J. Zaanen, T. Kondo, T. Takeuchi, J. R. Cooper, C. Putzke, S. Friedemann, A. Carrington,  and  N. E. Hussey, Incoherent transport across the strange-metal regime of overdoped cuprates. Nature, {\bf 595}, 661 (2021).

\bibitem{pepin2005} C. P\'epin, Fractionalization and Fermi-surface volume in heavy-fermion compounds:
 The case of YbRh$_2$Si$_2$. Phys. Rev. Lett. {\bf 94}, 066402 (2005).



\bibitem{Arovas1988} D. P. Arovas and A. Auerbach, Functional integral theories of low-dimensional quantum Heisenberg models. Phys. Rev. B {\bf 38}, 316 (1988).

\bibitem{parcollet1997transition} O. Parcollet and A. Georges,  Transition from overscreening to underscreening in the multichannel Kondo model: exact solution at large $N$. Phys. Rev. Lett. {\bf 79}, 4665 (1997).

\bibitem{Read1992Kagome} S. Sachdev, Kagom\'e- and triangular-lattice Heisenberg antiferromagnets: Ordering from quantum fiuctuations and quantum-disordered ground states with unconfined bosonic spinons. Phys. Rev. B {\bf 45}, 12377 (1992).

\bibitem{FlintSpN2009} R. Flint and P. Coleman, Symplectic $N$ and time reversal in frustrated magnetism. Phys. Rev. B {\bf 79}, 014424 (2009).

\bibitem{WangFa2006} F. Wang and A. Vishwanath, Spin-liquid states on the triangular and Kagom\'e lattices: a projective-symmetry-group analysis of Schwinger boson states. Phys. Rev. B {\bf 74}, 174423 (2006).

\bibitem{Wang2022_3IK} J. Wang and Y.-F. Yang, Spin current Kondo effect in frustrated Kondo systems. Sci. China-Phys. Mech. Astron. {\bf 65}, 227212 (2022).

\bibitem{Coleman-SWB-2006} J. Rech, P. Coleman, G. Zarand, and O. Parcollet, Schwinger boson approach to the fully screened Kondo model. Phys. Rev. Lett.  {\bf 96}, 016601 (2006). 

\bibitem{Yashar1D} Y. Komijani and P. Coleman, Model for a ferromagnetic quantum critical point in a 1D Kondo Lattice. Phys. Rev. Lett.  {\bf 120}, 157206 (2018).
 
\bibitem{Komijani2019} Y. Komijani and P. Coleman, Emergent critical charge fluctuations at the Kondo breakdown of heavy fermions. Phys. Rev. Lett.  {\bf 122}, 217001 (2019). 

\bibitem{Wang2019quantum} J. Wang, Y.-Y. Chang, C.-Y. Mou, S. Kirchner, and C.-H. Chung, Quantum phase transition in a two-dimensional Kondo-Heisenberg model: A dynamical Schwinger-boson large-$N$ approach. Phys. Rev. B {\bf 102}, 115133 (2020). 



\bibitem{Wang2022_FM} J. Wang and Y.-F. Yang, A unified theory of ferromagnetic quantum phase transitions in heavy fermion metals. Sci. China-Phys. Mech. Astron. {\bf 65}, 257211 (2022).

\bibitem{Han2021} R. Han, D. Hu, J. Wang, and Y.-F. Yang,  Schwinger boson approach for the dynamical mean-field theory of the Kondo lattice. Phys. Rev. B {\bf 104}, 245132 (2021).

\bibitem{supp} See Supplemental Material for more details.

\bibitem{coleman2005sum} P. Coleman, I. Paul, and J. Rech, Sum rules and Ward identities in the Kondo lattice. Phys. Rev. B {\bf 72}, 094430 (2005).

\bibitem{LohneysenRMP2007} H. v. L$\ddot{\text{o}}$hneysen, A. Rosch, M. Vojta, and P. W$\ddot{\text{o}}$lfle, Fermi-liquid instabilities at magnetic quantum phase transitions. Rev. Mod. Phys. {\bf 79}, 1015 (2007).


\bibitem{Kogut1979RMP} J. B. Kogut, An introduction of lattice gauge theory and spin systems. Rev. Mod. Phys. {\bf 51}, 659 (1979).

\bibitem{Sachdev2011vison} Y. Huh, M. Punk, and S. Sachdev, Vison states and confinement transitions of $\mathbb{Z}_2$ spin liquids on the kagome lattice. Phys. Rev. B {\bf 84}, 094419 (2011).

\bibitem{Moessner2001} R. Moessner, S. L. Sondhi, and E. Fradkin, Short-ranged resonating valence bond physics, quantum dimer models, and Ising gauge theories. Phys. Rev. B {\bf 65}, 024504 (2001).

\bibitem{Gazit2017} S. Gazit, M. Randeria, and A. Vishwanath, Emergent Dirac fermions and broken symmetries in confined and deconfined phases of Z$_2$ gauge theories. Nat. Phys. {\bf 13}, 484 (2017).

\bibitem{Motome2010PKS} Y. Motome, K. Nakamikawa, Y. Yamaji, and M. Udagawa, Partial Kondo screening in frustrated Kondo lattice systems. Phys. Rev. Lett. {\bf 105}, 036403 (2010).

\bibitem{Assaad2015DMFT} M. W. Aulbach, F. F. Assaad, and M. Potthoff, Dynamical mean-field study of partial Kondo screening in the periodic Anderson model on the triangular lattice. Phys. Rev. B {\bf 92}, 235131 (2015).




\end{thebibliography}
\end{document}


\title{$\mathbb{Z}_2$ metallic spin liquid on a frustrated Kondo lattice\\
\vspace{0.2cm}
- Supplemental Material -}
\author{Jiangfan Wang}
\affiliation{Beijing National Laboratory for Condensed Matter Physics and Institute of Physics,
Chinese Academy of Sciences, Beijing 100190, China}
\author{Yi-feng Yang}
\email[]{yifeng@iphy.ac.cn}
\affiliation{Beijing National Laboratory for Condensed Matter Physics,  Institute of Physics, 
Chinese Academy of Sciences, Beijing 100190, China}
\affiliation{School of Physical Sciences, University of Chinese Academy of Sciences, Beijing 100190, China}
\affiliation{Songshan Lake Materials Laboratory, Dongguan, Guangdong 523808, China}

\maketitle

In this supplemental material, we provide some important derivations in detail and additional figures that may be helpful for readers.

\subsection{I. Large-$N$ self-consistent equations}

To derive the large-$N$ action of the triangular lattice  Kondo-Heisenberg model, we first use the SU(2) Schwinger boson representation to rewrite the Kondo and Heisenberg terms as follows:
\begin{eqnarray}
H_K&=&J_K\sum_{i}\bm{S}_{i}\cdot \bm{s}_i=-\frac{J_K}{2}\sum_{ia}b_{i\alpha}^\dagger c_{i\alpha a}c_{i\beta a}^\dagger b_{i\beta}-\frac{J_KS}{2}\sum_{ia}c_{i\alpha a}^\dagger c_{i\alpha a},\notag \\
H_H&=&J_H\sum_{\langle ij\rangle}\bm{S}_i\cdot \bm{S}_j=\frac{J_H}{4}\sum_{\langle ij\rangle}\left(b_{i\alpha}^\dagger b_{j\alpha}b_{j\beta}^\dagger b_{i\beta}-\tilde{\alpha}b_{j\alpha}^\dagger b_{i,-\alpha}^\dagger\tilde{\beta}b_{j\beta}b_{i,-\beta}\right),
\end{eqnarray}
where $\tilde{\alpha}\equiv \text{sgn}(\alpha)$, and some unimportant constants have been dropped. The second term of $H_K$ can be absorbed into the definition of  the electron chemical potential. Upon Hubbard-Stratonovich transformations, the above interaction terms can be decomposed as:
\begin{eqnarray}
H_K&\rightarrow & \sum_{ia}\left(b_{i\alpha}^\dagger c_{i\alpha a}\chi_{ia}+c.c. +\frac{2}{J_K}|\chi_{ia}|^2\right),\notag \\
H_H&\rightarrow & \sum_{\langle ij\rangle }\left(\sum_{\alpha} (\tilde{\alpha}b_{j\alpha}^\dagger b_{i,-\alpha}^\dagger \Delta_{ji}-b_{i\alpha}^\dagger b_{j\alpha}\Gamma_{ji})+c.c.+\frac{4}{J_H}(|\Delta_{ij}|^2-|\Gamma_{ij}|^2)\right).
\end{eqnarray}
We then extend the number of spin flavor from 2 to $N$, and rescale the coupling constants  and the holon field through $J_K\rightarrow \frac{2}{N}J_K$, $J_H\rightarrow \frac{2}{N}J_H$, $\chi_{ia}\rightarrow \frac{1}{\sqrt{N}}\chi_{ia}$, in order to allow for a large-$N$ expansion. One then obtains a Sp($N$) invariant Lagrangian, $\mathcal{L}=\mathcal{L}_c+\mathcal{L}_\chi+\mathcal{L}_b+\mathcal{L}_{\text{int}}$, with
\begin{eqnarray}
\mathcal{L}_c &=&\sum_{{\bf p}\alpha a}c_{{\bf p}\alpha a}^\dagger (\partial_\tau+\epsilon_{\bf p})c_{{\bf p}\alpha a},\qquad \mathcal{L}_\chi =\sum_{ia}\frac{|\chi_{ia}|^2}{J_K}, \qquad \mathcal{L}_{\text{int}}=\frac{1}{\sqrt{N}}\sum_{i\alpha a}\left(b_{i\alpha}^\dagger c_{ia\alpha}\chi_{ia}+c.c.\right), \notag \\
\mathcal{L}_b &=&\sum_{i \alpha }b_{i\alpha}^\dagger(\partial_\tau+\lambda_i)b_{i\alpha}+\sum_{\langle ij\rangle \alpha}\left(\tilde{\alpha}b_{j\alpha}^\dagger b_{i,-\alpha}^\dagger \Delta_{ji}-b_{i\alpha}^\dagger b_{j\alpha}\Gamma_{ji}+c.c.\right)\notag \\
&&-2S\sum_i\lambda_i +\frac{2N}{J_H}\sum_{\langle ij\rangle}(|\Delta_{ij}|^2-|\Gamma_{ij}|^2).
\end{eqnarray} 
The partition function
\begin{equation}
\mathcal{Z}=\int \mathcal{D}[c,b,\chi,\lambda,\Delta,\Gamma]\exp\left(-\int_0^\beta d\tau \mathcal{L}\right)
\end{equation}
faithfully represents the original Kondo-Heisenberg model if all the functional integrals are performed exactly. 

At large $N$, one can replace the functional integrals over $\lambda_i$, $\Delta_{ij}$ and $\Gamma_{ij}$ by their saddle point values. One then obtains the following action in the Fourier space: 
\begin{eqnarray}
S &=&-\sum_{p \alpha a}( i\omega_n-\epsilon _{\bf p})c _{p \alpha a}^\dagger c _{p \alpha a}-\sum_{p \alpha }( i\nu_n -\varepsilon_{\bf p} ) b_{p \alpha }^{\dagger}b_{p \alpha }-i\sum_{p \alpha }\tilde{\alpha}b_{p\alpha}b_{-p,-\alpha }h_{\mathbf{p}}+c.c.  \notag \\
&&+\frac{1}{\sqrt{\beta\mathcal{V}N}}\sum_{pk \alpha a}b_{p \alpha }^{\dagger}c_{k \alpha a}\chi_{p-k, a}+c.c.+\sum_{p a}\frac{\left\vert \chi_{pa}\right\vert ^{2}}{J_{K}}+S_0, \label{eq:action}
\end{eqnarray}
where $S_0=N\beta \mathcal{V}\left(6(\Delta_0^2-\Gamma_0^2)/J_H-\lambda_0 \kappa \right)$, and we have used the simplified notation $p\equiv ({\bf p}, i\omega_n)$. The bare dispersions on the triangular lattice are 
$\epsilon_{\bf p}=2t\tilde{\varepsilon}_{\bf p}-\mu$, $\varepsilon_{\bf p}=\lambda_0-2\Gamma_0\tilde{\varepsilon}_{\bf p}$ and $h_{\bf p}=\Delta_0 \tilde{h}_{\bf p}$, with $\tilde{\varepsilon}_{\bf p}=\cos p_x+2\cos \frac{\sqrt{3}p_y}{2}\cos \frac{p_x}{2}$ and $\tilde{h}_{\bf p}=\sin p_x-2\cos \frac{\sqrt{3}p_y}{2}\sin \frac{p_x}{2}$.  For simplicity, we choose $t=-1/6$ and $\mu=0$ so that the electron band locates within the energy range $[-1,1/2]$.

The full Green's functions for spinons and holons are
\begin{eqnarray}
G_b({\bf p}, i\nu_n)&=&\frac{\gamma_b({\bf -p},-i\nu_n)}{\gamma_b({\bf p},i\nu_n)\gamma_b({\bf -p},-i\nu_n)-4h_{\bf p}^2}, \notag \\
G_\chi({\bf p},i\omega_n)&=&\frac{1}{-J_K^{-1}-\Sigma_\chi({\bf p},i\omega_n)}, \label{eq: GG}
\end{eqnarray}
where $\gamma_b({\bf p},i\nu_n)\equiv i\nu_n-\varepsilon_{\bf p}-\Sigma_b({\bf p}, i\nu_n)$, and the self-energies are given by Eq. (3) of the main text. These equations can be derived  rigorously from the Luttinger-Ward functional \cite{coleman2005sum} or the Dyson-Schwinger equations \cite{Wang2019quantum} at large $N$. 

The Green's functions are solved under three constraints corresponding to the minimization equations $\partial F/\partial \lambda_0=\partial F/\partial \Delta_0=\partial F/\partial \Gamma_0=0$, where $F$ is the free energy. These constraints are:
\begin{eqnarray}
\kappa&=&\frac{-1}{\beta\mathcal{V}}\sum_{{\bf p}n}G_b({\bf p},i\nu_n),\notag \\
\frac{3}{J_H}&=&\frac{1}{\beta\mathcal{V}}\sum_{{\bf p}n}\frac{\tilde{h}_{\bf p}^2}{\gamma_b({\bf p},i\nu_n)\gamma_b({\bf -p},-i\nu_n)-4h_{\bf p}^2},\notag \\
\frac{6\Gamma_0}{J_H}&=&\frac{1}{\beta\mathcal{V}}\sum_{{\bf p}n}\tilde{\varepsilon}_{\bf p}G_b({\bf p},i\nu_n). \label{eq: con}
\end{eqnarray}
In practice, we combine the last two equations of Eq. (\ref{eq: con}) to completely eliminate $J_H$ from the constraints. In this way, one can choose $\Delta_0$ as an input parameter and solve only two constraints to obtain $\lambda_0$ and $\Gamma_0$. The value of $J_H$ should be determined by Eq. (\ref{eq: con}) after  self-consistencies are achieved. However, naively using $J_H$ as a tuning parameter may lead to artificial first-order transitions, a known pathology of the Schwinger boson large-$N$ theory \cite{fczhang2002pathology}. Fortunately, this problem can be cured by adding an extra biquadratic spin interaction term, $-\zeta J_H\sum_{\langle ij \rangle}(\bm{S}_i\cdot \bm{S}_j)^2$, to the original Heisenberg Hamiltonian \cite{Yashar1D,Wang2021nonlocal}, which amounts to modify the Heisenberg coupling and turn the artificial first-order transitions into continuous transitions.

\subsection{II. Heisenberg limit}

In the Heisenberg limit ($J_K=0$), the electrons completely decouple with the spinons, so that the spinon self-energy vanishes. The spinon Green's function then reduces to
\begin{equation}
G_{b0}({\bf p}, i\nu_n)=\frac{-i\nu_n-\varepsilon_{\bf p}}{(i\nu_n-\varepsilon_{\bf p})(-i\nu_n-\varepsilon_{\bf p})-4h_{\bf p}^2},
\end{equation}
which gives the bare spinon dispersion $E_{\bf p}^s=\sqrt{\varepsilon_{\bf p}^2-4h_{\bf p}^2}$. The constraints are simplified to the following equations:
\begin{eqnarray}
2\kappa&=&\frac{1}{\mathcal{V}}\sum_{\bf p}\left(\frac{\varepsilon_{\bf p}}{E_{\bf p}^s}\coth \frac{\beta E_{\bf p}^s}{2}-1\right),\notag \\
\frac{6}{J_H}&=&\frac{1}{\mathcal{V}}\sum_{\bf p}\frac{\tilde{h}_{\bf p}^2}{E_{\bf p}^s}\coth \frac{\beta E_{\bf p}^s}{2},\notag \\
\frac{12\Gamma_0}{J_H}&=&\frac{-1}{\mathcal{V}}\sum_{\bf p}\tilde{\varepsilon}_{\bf p}\left(\frac{\varepsilon_{\bf p}}{E_{\bf p}^s}\coth \frac{\beta E_{\bf p}^s}{2}-1\right). \label{eq:ConH}
\end{eqnarray}

\begin{figure}
\centering\includegraphics[scale=0.6]{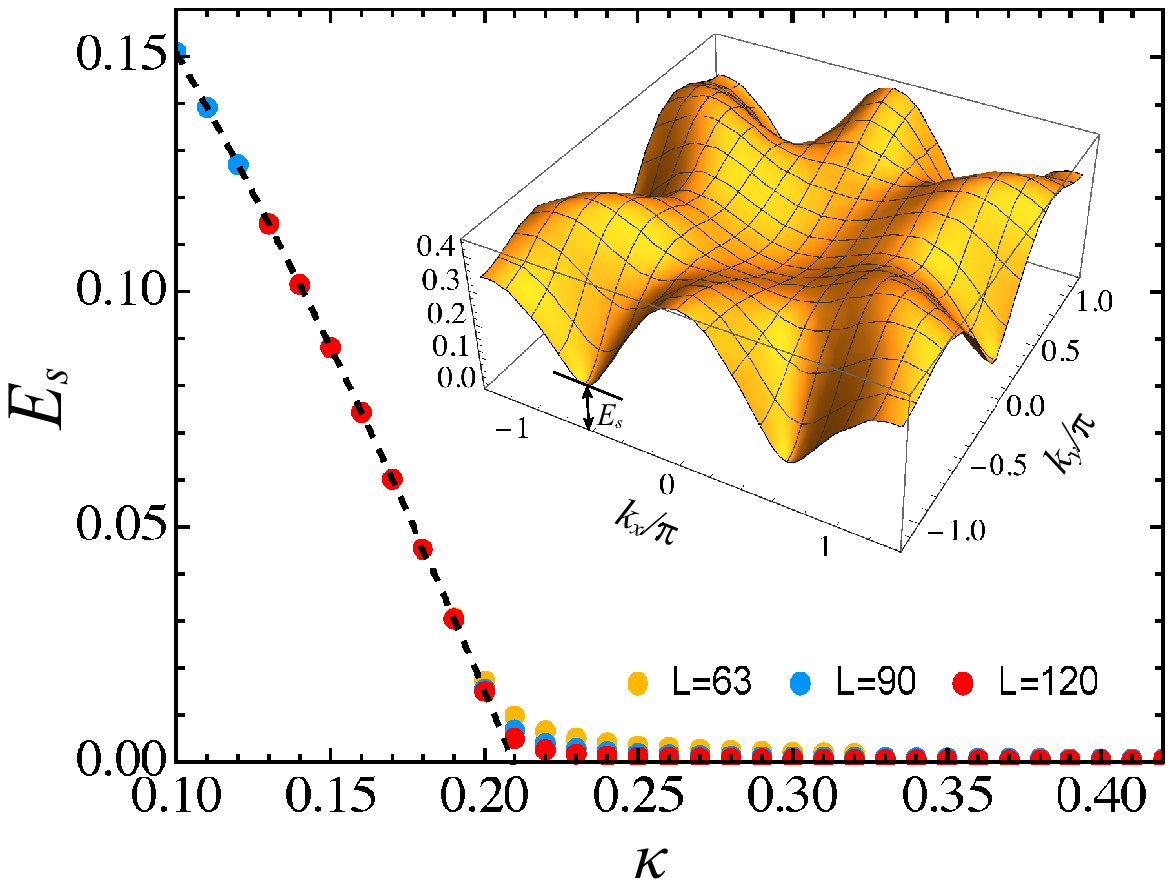}
\caption{The spinon gap ($E_s$) as a function of $\kappa$ in the Heisenberg limit ($J_K=0$). $L=63$, $90$ and $120$ are the lattice sizes. The dashed line is a fitting of the gap in the paramagnetic phase, which extrapolates to the critical point $\kappa_c=0.21$ for infinite lattice size. The inset shows the spinon dispersion at $\kappa=0.13$.}
\label{fig:Heisenb}
\end{figure}

For fixed $J_H=1$ and different values of $\kappa$, one can then solve Eq. (\ref{eq:ConH}) to obtain the spinon dispersion. As shown in Fig. \ref{fig:Heisenb}, the spinon has a gap at small $\kappa$, which decreases monotonically as increasing $\kappa$, and eventually becomes gapless at some critical value $\kappa_c$, leading to a quantum phase transition from a gapped spin liquid to an antiferromagnetic order through the spinon condensation. Our calculation gives a critical $\kappa_c=0.21$,  larger than the $\kappa_c=0.17$ obtained in Ref. \cite{Read1992Kagome}, which used only  one auxiliary field  $\Delta_{ij}$. Using two auxiliary fields has been shown to describe the frustrated spin systems better \cite{FlintSpN2009}. In fact, it is physically reasonable that $\kappa_c$ of the triangular lattice is larger than that of the square lattice ($\kappa_c=0.197$) due to  geometric frustrations.

\subsection{III. Numerical methods}

\begin{figure}
\centering\includegraphics[scale=0.22]{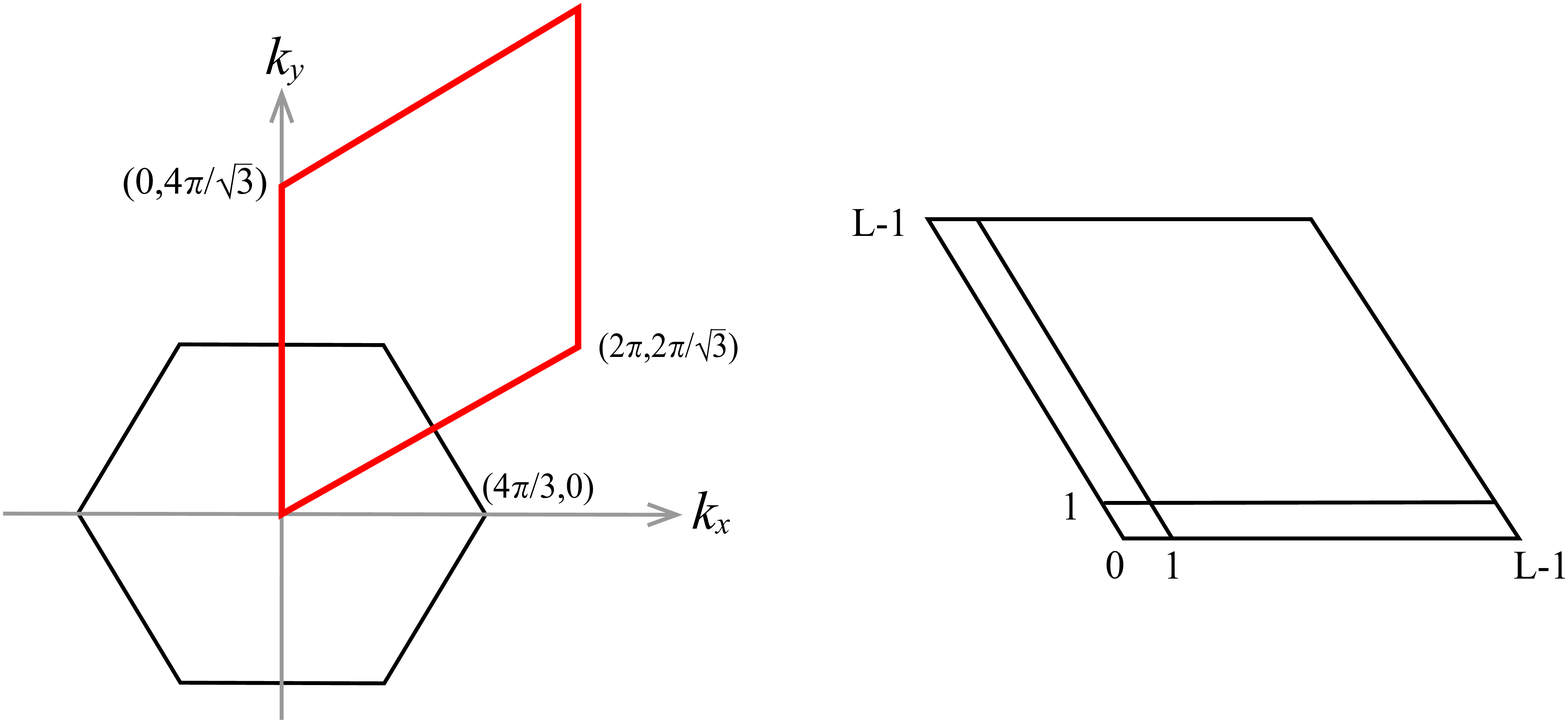}
\caption{The Brillouin zone (red) and the real space sample (right) used to perform the fast Fourier transform.}
\label{fig:Lattice}
\end{figure}

To perform numerical calculations, we choose the diamond-shape Brillouin zone instead of the hexagonal one as shown in Fig. \ref{fig:Lattice}. The Brillouin zone is discretized as:
\begin{equation}
k_x=\frac{2\pi n}{L},\qquad k_y=\frac{2\pi(2m+n)}{\sqrt{3}L},\qquad n,m=0,\cdots,L-1
\end{equation}
where $L$ (chosen as 72 in our calculations) is the sample size. Since the lattice coordination in real space is specified by $\bm{r}=\left(i-\frac{j}{2},\frac{\sqrt{3}j}{2}\right)$ with $i,j=0,\cdots,L-1$, we can write the Fourier transform of any function $G(\bm{r})$ as follows:
\begin{eqnarray}
G(\bm{r})&=&\frac{1}{V_{\text{B.Z.}}}\int_{{\bf k}\in \text{B.Z.}}G({\bf k})e^{i{\bm k}\cdot {\bm r}}\notag \\
&=&\frac{1}{8\pi^2/\sqrt{3}}\frac{8\pi^2}{\sqrt{3}L^2}\sum_{n,m=0}^{L-1}G(n,m)e^{i\left[\frac{2\pi n}{L}(i-\frac{j}{2})+\frac{2\pi (2m+n)}{\sqrt{3}L}\frac{\sqrt{3}j}{2}\right]}\notag\\
&=&\frac{1}{L^2}\sum_{n,m=0}^{L-1}G(n,m)e^{i\left[\frac{2\pi n}{L}i+\frac{2\pi m}{L}j\right]}=G(i,j). \label{eq:FFT}
\end{eqnarray}
The inverse Fourier transform is defined as
\begin{equation}
G(n,m)=\sum_{i,j=0}^{L-1}G(i,j)e^{-i\left[\frac{2\pi n}{L}i+\frac{2\pi m}{L}j\right]}. \label{eq:FFT2}
\end{equation}
Eqs. (\ref{eq:FFT}) and (\ref{eq:FFT2}) can be performed efficiently using the fast Fourier transform (FFT) algorithm. 

To solve the self-consistent equations, we first write the self-energy equations (Eq. (3) of the main text) into the real frequency space,
\begin{eqnarray}
\Sigma_b({\bf p},\omega)&=&\frac{-\kappa}{\mathcal{V}}\sum_{\bf k}\int\frac{dz}{\pi}n_F(z)\left[g_c({\bf p-k},z+\omega)\text{Im}G_\chi({\bf k},-z)-G_\chi({\bf k}, \omega-z)\text{Im}g_c({\bf p-k}, z)\right], \notag\\
\Sigma_\chi({\bf p},\omega)&=&\frac{1}{\mathcal{V}}\sum_{\bf k}\int\frac{dz}{\pi}\left[n_B(z)g_c({\bf k-p},z-\omega)^*\text{Im}G_b({\bf k},z)-n_F(z)G_b({\bf k}, \omega+z)\text{Im}g_c({\bf k-p}, z)\right],\label{eq:SelfE}
\end{eqnarray}
where $n_B(z)$ and $n_F(z)$ are the bosonic and fermionic distribution functions. After Fourier transform, the momentum convolutions in Eq. (\ref{eq:SelfE}) become direct multiplications in the coordinate space,
\begin{eqnarray}
\Sigma_b({\bf r},\omega)&=&-\kappa\int\frac{dz}{\pi}n_F(z)\left[g_c({\bf r},z+\omega)\text{Im}G_\chi({\bf r},-z)-G_\chi({\bf r}, \omega-z)\text{Im}g_c({\bf r}, z)\right], \notag\\
\Sigma_\chi({\bf r},\omega)&=&\int\frac{dz}{\pi}\left[n_B(z)g_c({\bf -r},z-\omega)^*\text{Im}G_b({\bf r},z)-n_F(z)G_b({\bf r}, \omega+z)\text{Im}g_c({\bf -r}, z)\right]. \label{eq:SelfE1}
\end{eqnarray}
In each iteration step, we first use FFT to transform the spinon and holon Green's functions obtained from the last step to the real space, then substitute them into Eq. (\ref{eq:SelfE1}) to calculate $\Sigma_b({\bf r},\omega)$ and $\Sigma_\chi({\bf r},\omega)$, and finally use the inverse FFT to transform them back to the momentum space. The lattice symmetry can be utilized to further reduce the computational efforts.

\subsection{IV. Spinon and holon spectra}

\begin{figure}
\centering\includegraphics[scale=0.47]{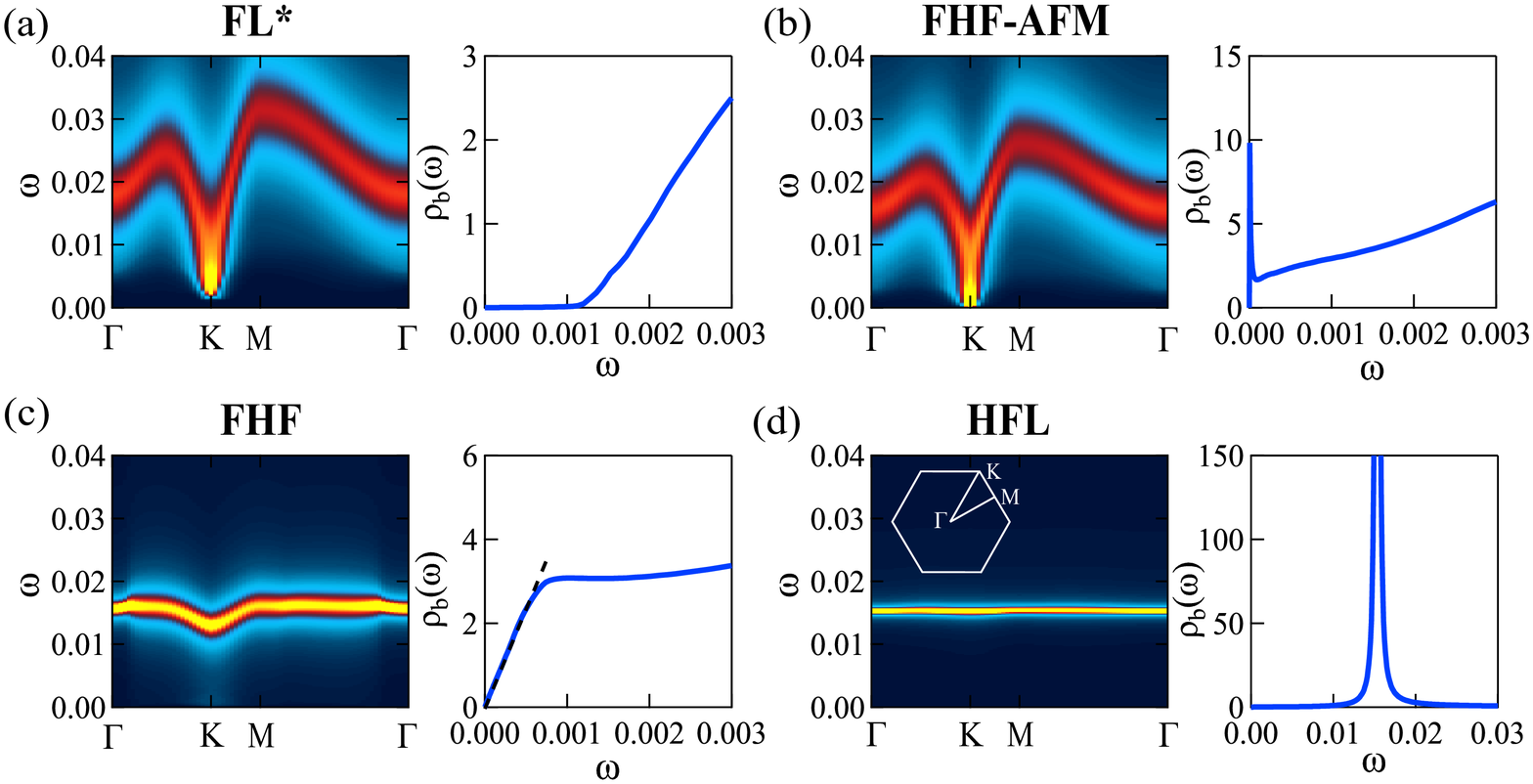}
\caption{The spinon spectra and its density of states for $\kappa=0.1$ and different values of $T_K/J_H$: (a) $0.123$ (FL$^*$); (b) $0.153$ (FHF-AFM QCP); (c) $0.27$ (FHF); (d) $0.33$ (HFL).}
\label{fig:Specb}
\end{figure}

Here we provide additional information for the spinon and holon spectra. As shown in Fig. \ref{fig:Specb} for  $\kappa=0.1$ and different values of $T_K/J_H$, the spinons are gapped in the FL$^*$ and the HFL states, but become gapless within the FHF state where $\rho_b(\omega)$ exhibits a linear-in-$\omega$ behavior at low energy, indicating highly damped spinons from scattering with electrons and holons. At the FHF-AFM QCP, $\rho_b(\omega)$ develops a $\delta$-like peak at zero energy due to the spinon condensation at the $K$ point, as shown in Fig. \ref{fig:Specb}(b). 

\begin{figure}
\centering\includegraphics[scale=0.49]{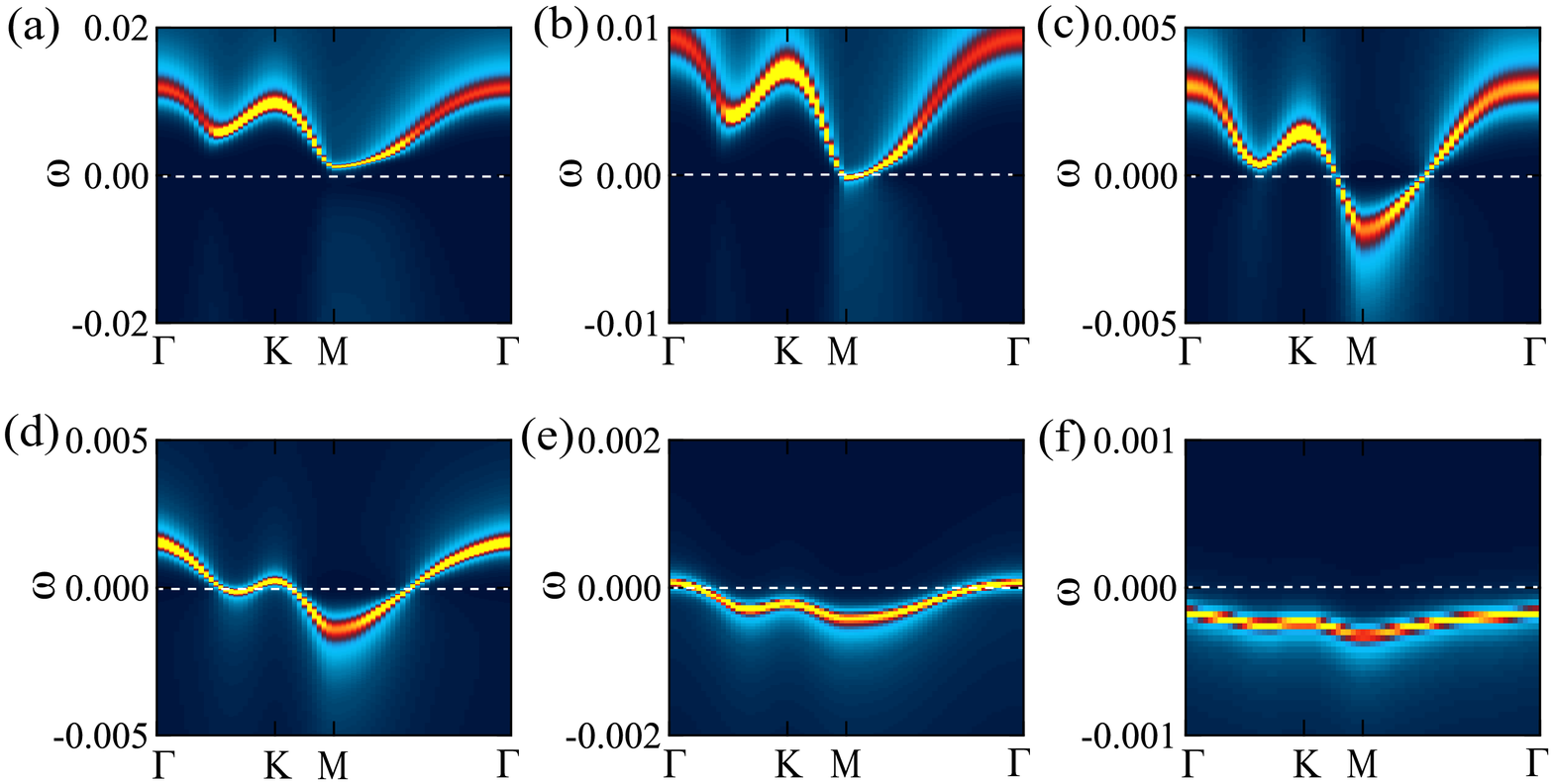}
\caption{The holon spectra for $\kappa=0.1$ and different values of $T_K/J_H$: (a) $0.123$ (FL$^*$); (b) $0.138$; (c) $0.206$; (d) $0.245$; (e) $0.304$; (f) $0.33$ (HFL). The white dashed line denotes the Fermi level. The corresponding holon Fermi surfaces of (b)-(e) are shown in Fig. 1(d) of the main text. }
\label{fig:Specx}
\end{figure}

Figure \ref{fig:Specx} shows the evolution of holon band from above to below the Fermi energy upon increasing $T_K/J_H$, a general feature of the Kondo lattice. The band is empty in the FL$^*$ state and fully occupied in the HFL state. In between, as the holon band crosses the Fermi level, its Fermi surface first emerges at the $M$ point, then gradually expands to the entire Brillouin zone, and eventually vanishes at the $\Gamma$ point. Approaching the FHF-HFL boundary, the band becomes extremely narrow, indicating an increasingly heavy effective mass of the holons. The bands in Figs. \ref{fig:Specx}(b)-(e) give rise to the holon Fermi surfaces shown in Fig. 1(d) of the main text. The partial filling of the holon band distinguishes the FHF state from the FL$^*$ and HFL states.

\subsection{V. Derivation of the $\mathbb{Z}_2$ gauge theory}

To consider the $\mathbb{Z}_2$ gauge fluctuations above the mean-field solutions,  we substitute $\Delta_{ij}=\Delta_0 Z_{ij}$ and $\Gamma_{ij}=\Gamma_0 Z_{ij}$ into the original Lagrangian,
\begin{eqnarray}
\mathcal{L}&=&\sum_{{\bf k}\alpha a}c_{{\bf k}\alpha a}^\dagger (\partial_\tau+\epsilon_{\bf k})c_{{\bf k}\alpha a}+\sum_{i \alpha }b_{i\alpha}^\dagger(\partial_\tau+\lambda_0)b_{i\alpha}+\sum_{ia}\frac{|\chi_{ia}|^2}{J_K}-2S\mathcal{V}\lambda_0 +\frac{6N\mathcal{V}}{J_H}(\Delta_0^2-\Gamma_0^2) \notag \\
&&+\frac{1}{\sqrt{N}}\sum_{i\alpha a}b_{i\alpha}^\dagger c_{ia\alpha}\chi_{ia}+c.c. +\sum_{\langle ij\rangle \alpha}Z_{ji}\left(\tilde{\alpha}b_{j\alpha}^\dagger b_{i,-\alpha}^\dagger \Delta_0-b_{i\alpha}^\dagger b_{j\alpha}\Gamma_0+c.c.\right). \label{eq:L}
\end{eqnarray}
In principle, there is also a time component  $\mathbb{Z}_2$ gauge field associated with the fluctuation of $\lambda_i$, corresponding to the term $b_{i\alpha}^\dagger(\tau_{m+1}) Z_{i}^{m,m+1}b_{i\alpha}(\tau_m)$, where $\tau_m$ is the discretized imaginary time. However, one can always choose a gauge so that $Z_i^{m,m+1}=1$ for all $i$ and $m$. Our goal is to obtain an effective theory describing the holons and $\mathbb{Z}_2$ gauge field. For this purpose, we integrate out both spinons and conduction electrons in the  partition function,
\begin{eqnarray}
\mathcal{Z}=\int \mathcal{D}[c,b,\chi,Z]\exp \left(-\int_0^\beta d\tau \mathcal{L}\right)\propto\int \mathcal{D}[\chi, Z]\exp \left(-S_{\text{eff}}\right).
\end{eqnarray}
The integration can be done perturbatively by treating the second line of Eq. (\ref{eq:L}) as interacting vertices with small parameters $1/\sqrt{N}$, $\Delta_0$ and $\Gamma_0$. The leading order terms of $S_{\text{eff}}$ is generated by the Feynman diagrams listed in Fig. 3 of the main text. As an example, here we compute the first diagram of Fig. 3, which involves the vertex $S_{\Delta}=\frac{\Delta_0}{\sqrt{\beta}}\sum_{\langle ij\rangle nl\alpha }\tilde{\alpha}b_{j\alpha}^\dagger (\nu_n) b_{i,-\alpha}^\dagger(\nu_l-\nu_n) Z_{ji}(\nu_l)$ and its complex conjugate. We have
\begin{eqnarray}
\langle S_\Delta \bar{S}_\Delta\rangle &=&\frac{2\Delta_0^2}{\beta}\sum_{\langle ij\rangle nl\alpha }Z_{ji}(\nu_l)Z_{ji}(-\nu_l)\langle b_{i,-\alpha}(\nu_l-\nu_n)b_{i,-\alpha}^\dagger(\nu_l-\nu_n)\rangle \langle b_{j\alpha}(\nu_n)b_{j\alpha}^\dagger (\nu_n)\rangle \notag\\
&=&2N\Delta_0^2\sum_{\langle ij\rangle l}Z_{ji}(\nu_l)Z_{ji}(-\nu_l)\frac{1}{\beta}\sum_{n}\frac{1}{i\nu_l-i\nu_n-\lambda_0}\frac{1}{i\nu_n-\lambda_0} \notag\\
&=&2N\Delta_0^2\sum_{\langle ij\rangle l}Z_{ji}(\nu_l)Z_{ji}(-\nu_l)\frac{1+2n_B(\lambda_0)}{2\lambda_0-i\nu_l},
\end{eqnarray}
where $n_B(\lambda_0)=0$ at zero temperature since $\lambda_0>0$. The long time dynamics can be obtained by a small frequency expansion, $\frac{1}{2\lambda_0-i\nu_l}\approx\frac{1}{2\lambda_0}\left(1+\frac{i\nu_l}{2\lambda_0}+\frac{i\nu_l^2}{4\lambda_0^2}\right)$, which leads to 
\begin{eqnarray}
\langle S_\Delta \bar{S}_\Delta\rangle &=&\frac{N\Delta_0^2}{\lambda_0}\sum_{\langle ij\rangle}\int_0^\beta d\tau \left( Z_{ji}^2 - \frac{1}{4\lambda_0^2}(\partial_\tau Z_{ji})^2\right).
\end{eqnarray}
Note that $\int_0^\beta d\tau Z_{ji}\partial_\tau Z_{ji}=-\int_0^\beta d\tau (\partial_\tau Z_{ji})Z_{ji}=0$ vanishes, and $Z_{ji}^2=1$.

The final effective action is obtained as
\begin{eqnarray}
S_{\text{eff}}&=&\int_0^\beta d\tau \left[\xi \sum_{\langle ij\rangle}(\partial_\tau Z_{ji})^2+\sum_{ia}\chi_{ia}^\dagger (\partial_\tau+\epsilon_\chi)\chi_{ia}+\bar{t}\sum_{\langle ij\rangle a}Z_{ji}\chi_{ia}^\dagger \chi_{ja}\right.\notag\\
&&\qquad \qquad\left.-K\sum_{(ijk)\in \triangle }Z_{ij}Z_{jk}Z_{ki}-K'\sum_{(ijkl)\in \lozenge}Z_{ij}Z_{jk}Z_{kl}Z_{li}+\cdots \right],  \label{eq:Seff}
\end{eqnarray}
where
\begin{eqnarray}
\xi&=&\frac{N\Delta_0^2}{4\lambda_0^3},\qquad \epsilon_\chi=\frac{1}{Z_\chi}\left(\frac{1}{J_K}-\frac{1}{\mathcal{V}}\sum_{\bf k}\frac{n_F(\epsilon_{\bf k})}{\lambda_0-\epsilon_{\bf k}}\right),\qquad Z_\chi=\frac{1}{\mathcal{V}}\sum_{\bf k}\frac{n_F(\epsilon_{\bf k})}{(\epsilon_{\bf k}-\lambda_0)^2}, \notag \\
\bar{t}&=&-\frac{2\Gamma_0}{Z_\chi \mathcal{V}}\sum_{\bf k}\frac{n_F(\epsilon_{\bf k})e^{i{\bf k}\cdot ({\bf r}_i-{\bf r}_j)}}{(\epsilon_{\bf k}-\lambda_0)^2},\quad K=-2N\frac{\Gamma_0\Delta_0^2}{\lambda_0^2},\quad K'=N\frac{1}{\lambda_0^3}\Delta_0^2(\Delta_0^2-4\Gamma_0^2),
\end{eqnarray}
and the holon field has been scaled by its quasiparticle residue as $\chi_{ia}\rightarrow \chi_{ia}/\sqrt{Z_\chi}$. In our calculations, the typical values of $\Delta_0$ and $\Gamma_0$ range from $10^{-3}\sim 10^{-5}$, while $\lambda_0\sim 10^{-2}$ does not vary too much as one changes $T_K/J_H$. For $N=2$, this leads to a broad range of $\xi$ from the order of unity to  $10^{-4}$, while $K$ and $K'$ are typically less than $10^{-5}$. Note that the integration over conduction electrons gives rise to momentum sums like $\sum_{\bf k}\frac{n_F(\epsilon_{\bf k})}{(\epsilon_{\bf k}-\lambda_0)^n}$, where $n$ is a positive integer. These sums are always finite since only negative $\epsilon_{\bf k}$ are involved at zero temperature while $\lambda_0>0$. For example, we have 
\begin{eqnarray}
\frac{1}{\mathcal{V}}\sum_{\bf k}\frac{n_F(\epsilon_{\bf k})}{\lambda_0-\epsilon_{\bf k}}= \int_{-D}^0d\epsilon \frac{\rho_c(\epsilon)}{\lambda_0-\epsilon}\approx\rho_c\ln\frac{D+\lambda_0}{\lambda_0}\approx \rho_c\ln \frac{D}{\lambda_0},
\end{eqnarray}
where we have used $D\gg \lambda_0$.

Since the continuous time derivative of a discrete field is not well defined, $\partial_\tau Z_{ji}$ must be viewed as a finite difference, $[Z_{ji}(\tau_m+\epsilon)-Z_{ji}(\tau_m)]/\epsilon$. The discrete time slice $\epsilon$ corresponds to the least time required for $\Delta_{ij}$ to tunnel from one minimum ($\Delta_0$) to the other ($-\Delta_0$), and is simply regarded as a high energy cutoff of our effective action. For small $\epsilon$, the dynamical term $\xi(\partial_\tau Z_{ji})^2$ is equivalent to the Hamiltonian $-g\hat{X}_{ji}$ with $g=\epsilon^{-1}\exp (-4\xi/\epsilon)$ \cite{Kogut1979RMP}, thus leading to the effective Hamiltonian Eq. (4) of the main text.